\documentclass[runningheads]{svmult}

\usepackage{makeidx}   
\usepackage{graphicx}  
\usepackage{subeqnar}  
\usepackage{multicol}  
\usepackage{physprbb}  
\makeindex             

\begin{document}
\title*{The black hole mass of BL Lacs 
from stellar velocity dispersion of the host galaxy.}
\toctitle{The black hole mass of BL Lacs 
\protect\newline from stellar velocity dispersion of the host galaxies}
%
%
\titlerunning{ The black hole mass of BL Lacs  }
%
\author{Renato Falomo\inst{1}
\and Jari Kotilainen\inst{2}
\and Aldo Treves\inst{3}
}
\authorrunning{Renato Falomo et al.}
%
%
\institute{Osservatorio Astronomico di Padova, Vicolo dell'Osservatorio 5, Padova, Italy
\and Tuorla Observatory, University of Turku, V\"ais\"al\"antie 20, Piikki\"o, Finland 
\and Universit\`a dell'Insubria, Via Valleggio 11, Como, Italy
}

\maketitle              

\section{Introduction}

One of the most important quantities in theoretical models of AGNs is the black
hole mass (M$_{BH}$) that, together with the total luminosity, defines 
the fraction of the Eddington luminosity at
which the AGN is emitting. Determination of M$_{BH}$ in AGN is 
difficult mainly because of the bright emission from the nucleus and their large
distance. The main method that has proved to be successful in AGN is
reverberation mapping, which is extremely time consuming and gives results on
M$_{BH}$ that depend on the assumed geometry of the accretion disk. Therefore,
only for a few well studied quasars and Seyfert galaxies M$_{BH}$ is known (see
e.g. [7], [11] and references therein). This method cannot
obviously be employed for BL Lac objects because they lack prominent
emission lines. Therefore other methods need to
be applied to infer M$_{BH}$  for BL Lacs. The discovery of a relation between
M$_{BH}$ and the luminosity of the bulge in nearby early-type galaxies offers
now a new tool for estimating the mass of the central BH (see e.g. review [10]). 
This has been done for two samples of nearby quasars [9] 
and BL Lacs [12].

Recently, a tighter correlation was found relating M$_{BH}$ with the central
stellar velocity dispersion $\sigma$ of the spheroidal component in nearby
galaxies [5,2], that can also be used to
estimate M$_{BH}$ in AGN. The relationship appears to predict more accurately
[10] M$_{BH}$, but requires the  measurement of $\sigma$ in the host 
galaxies of AGN
that is difficult to obtain, in particular for objects at moderately high
redshift and with very luminous nuclei. On the other hand, 
for BL Lacs that have 
relatively fainter nuclei than quasars, this measurement (at least for low 
redshift objects) can be secured with observations at medium-sized telescopes.

We present here the first estimates of stellar velocity dispersion of BL Lacs
from our ongoing program aimed specifically at deriving  M$_{BH}$ from the
M$_{BH}$  -- $\sigma$ correlation. We selected a sample of nearby (z$<$0.2) BL
Lacs  for which high quality images were obtained either from the ground 
using the Nordic Optical Telescope (NOT) [3]  or with HST+WFPC2 [13,4]. 
From the images a characterization of the host galaxies and of  
the nuclear luminosity are obtained.
This allows us to compare  M$_{BH}$ with the mass (and the luminosity) of the
host galaxy and also to evaluate the Eddington ratio, provided that the nuclear
emitted power is corrected for the beaming factor. Moreover, a comparison of
M$_{BH}$  for BL Lacs  with different  jet/ disk luminosities can be used to test
the hypothesis (see e.g. [8]) that the  accretion rate
changes from largely sub-Eddington, for low luminosity weak-lined sources, to
near-Eddington for high luminosity, strong-lined  sources. If the accretion rate
in terms of Eddington ratio were the same in both classes, the BH masses should
differ almost by three orders of magnitude.

\section{Observations, data analysis and first results}

We secured medium resolution (R $\sim$3000) optical spectra of a sample of the
BL Lacs using the NOT 2.5m and the ESO 3.6m telescopes equipped with ALFOSC and
EFOSC2, respectively. The chosen grisms combined with a 1 arcsec slit yield a
spectral resolution for velocity dispersion measurement of $\sim$60-80 km/s. The
used spectral range includes the absorption lines of Ca II (3933-68 \AA), Mg I
(5175 \AA), E-band (5269 \AA) and Na I (5892 \AA) from the host galaxies.
The measurement of $\sigma$ was done with the Fourier Quotient method using
template spectra of late-type (G and K) stars.

Here we report the first results on three BL Lacs: Mkn 501, Mkn 180 and PKS
2201+04. From the measurement of $\sigma$ we have evaluated the black hole
masses assuming that the M$_{BH}$ -- $\sigma$ relationship [2] is valid for the
BL Lac hosts. We found similar BH masses (log(M$_{BH}$) = 8.5 and 8.7) for Mkn
501 and Mkn 180, and a significantly lower value (log(M$_{BH}$) = 7.5) for PKS
2201+04.

From the values of $\sigma$ and the
effective radius R$_e$ we have also estimated the mass of the host galaxy (using
the relation M(host) = 5 R$_e$$\sigma^2$/G; [1] and compared it
with that of the BH. It turns out that M$_{BH}$/M(host) for the BL Lacs is in
the range of 0.03 - 0.1\% .

%
\end{document}